\documentclass[graybox]{svmult}

\usepackage{mathptmx}       % selects Times Roman as basic font
\usepackage{helvet}         % selects Helvetica as sans-serif font
\usepackage{courier}        % selects Courier as typewriter font
\usepackage{type1cm}        % activate if the above 3 fonts are
\usepackage{makeidx}         % allows index generation
\usepackage{graphicx}        % standard LaTeX graphics tool
\usepackage{multicol}        % used for the two-column index
\usepackage[bottom]{footmisc}% places footnotes at page bottom

\usepackage{newtxtext}
\usepackage[varvw]{newtxmath} % selects Times Roman as basic font

\usepackage[colorlinks=true,urlcolor=blue]{hyperref}
\usepackage{siunitx}

\DeclareSIUnit{\au}{au}
\DeclareSIUnit{\ME}{M_{\oplus}}
\DeclareSIUnit{\RM}{R_{M}}
\DeclareSIUnit{\MM}{M_{M}}
\DeclareSIUnit{\RJ}{R_{J}}
\DeclareSIUnit{\MJ}{M_{J}}

\makeindex             % used for the subject index
\begin{document}

\title*{Giant Planet Formation by Disk Instability}
% Use \titlerunning{Short Title} for an abbreviated version of
% your contribution title if the original one is too long
\author{Ravit Helled \orcidID{0000-0001-5555-2652} and
\\ Oliver Schib\orcidID{0000-0001-6693-7910} and 
\\Christian Reinhardt\orcidID{0000-0002-4535-3956} and 
\\Noah Kubli \orcidID{0009-0004-4569-6782} and 
\\Lucio Mayer \orcidID{0000-0002-7078-2074} and 
\\Christoph Mordasini \orcidID{ 0000-0002-1013-2811} and  
\\Gabriele Cugno \orcidID{0000-0001-7255-3251} }
% Use \authorrunning{Short Title} for an abbreviated version of
% your contribution title if the original one is too long

\institute{
Ravit Helled \at Department of Astrophysics, University of Zurich, Winterthurerstrasse 190, CH-8059 Zurich, Switzerland.
\email{ravit.helled@uzh.ch}
\and
Oliver Schib \at Division of Space Research and Planetary Sciences, Physics Institute, University of Bern, Gesellschaftsstrasse 6, CH-3012 Bern, Switzerland.
\email{oliver.schib@unibe.ch}
\and
Christian Reinhardt \at
Department of Astrophysics, University of Zurich, Winterthurerstrasse 190, CH-8059 Zurich, Switzerland.
\email{christian.reinhardt@uzh.ch}
\at
Division of Space Research and Planetary Sciences, Physics Institute, University of Bern, Gesellschaftsstrasse 6, CH-3012 Bern, Switzerland.
\email{christian.reinhardt@unibe.ch}
\and
Lucio Mayer \at Department of Astrophysics, University of Zurich, Winterthurerstrasse 190, CH-8059 Zurich, Switzerland.
\email{lucio.mayer@uzh.ch}
\and
Noah Kubli \at Department of Astrophysics, University of Zurich, Winterthurerstrasse 190, CH-8059 Zurich, Switzerland.
\email{noah.kubli@uzh.ch}
\and
Christoph Mordasini \at Division of Space Research and Planetary Sciences, Physics Institute, University of Bern, Gesellschaftsstrasse 6, CH-3012 Bern, Switzerland.
\email{christoph.mordasini@unibe.ch}
\and
Gabriele Cugno \at Department of Astrophysics, University of Zurich, Winterthurerstrasse 190, CH-8059 Zurich, Switzerland.
\email{gabriele.cugno@uzh.ch}
}
%
% Use the package "url.sty" to avoid
% problems with special characters
% used in your e-mail or web address
%
\maketitle

\abstract*{Each chapter should be preceded by an abstract (no more than 200 words) that summarizes the content. The abstract will appear \textit{online} at \url{www.SpringerLink.com} and be available with unrestricted access. This allows unregistered users to read the abstract as a teaser for the complete chapter.
Please use the 'starred' version of the \texttt{abstract} command for typesetting the text of the online abstracts (cf. source file of this chapter template \texttt{abstract}) and include them with the source files of your manuscript. Use the plain \texttt{abstract} command if the abstract is also to appear in the printed version of the book.}

\abstract{The disk instability (DI) model for giant planet formation remains an attractive alternative in explaining the formation of giant planets at early times, giant planets at large radial distances, and giant planets orbiting M-stars. 
In this review, we present recent developments in the disk instability model including hydrodynamical as well as magneto-hydrodynamical (MHD) disk simulations, populations synthesis models, and  simulations of clump-clump collisions. We also discuss advances in observations that can be used to constrain and test this formation scenario. }

\section{Overview}
The formation of planets is a fundamental process in astrophysics, underpinning our understanding of the origins and diversity of planetary systems.
Although the core accretion model \cite{Pollack1996,Alibert2005,YoudinZhu2025} has long been the dominant paradigm, the alternative mechanism, disk instability (DI), remains valid due to its ability to explain the rapid formation of massive planets in protoplanetary disks, the formation of giant planets around low-mass stars, and the formation of massive giant planets at large radial distances. Disk instability involves the gravitational fragmentation of a sufficiently massive and cool disk, leading to the direct formation of gas giant planets on short timescales \cite{Boss1997, Durisen2007, Helled2014}.

Recent observational advancements, including high-resolution imaging of protoplanetary disks with facilities like ALMA \cite{Andrews2020} and the upcoming capabilities of the Extremely Large Telescope (ELT), provide unprecedented opportunities to test the predictions of the DI model. These observations reveal intricate substructures within disks, such as spiral arms and clumps, which may be indicative of gravitational instabilities \cite{Kratter2016}. However, significant uncertainties remain regarding the conditions under which disk instability operates and its relative contribution to planet formation compared to core accretion.

In this review, we do not discuss the role of gravitational instability in the formation of binary systems and brown dwarfs (see \cite{Kratter2016,2015MNRAS.449.3432S,2014prpl.conf..619C} and references therein) or the role of DI in forming free-floating planets. Instead we summarize the current understanding of giant planet formation via disk instability and recent developments lead by Swiss researchers involved in the NCCR PlanetS. We emphasize the interplay between theoretical models, numerical simulations, and observational constraints. We also identify key open questions, such as the role of disk fragmentation on the evolution of disks and the planet formation  processes.  

\section{The basic physics of disk instability }
In the DI model, giant planets form when a gravitationally unstable region of the protoplanetary disk fragments under it own gravity to form bound gaseous clumps that may evolve to become giant planets. 
Disk fragmentation can result from density enhancements throughout
regions of a disk wherever the destabilizing effects of self-gravity overcome the stabilizing effects of pressure and shear.
This is typically the case in very young disks at tens of Astronomical Units (AU).
The growth of small local density perturbations in a
thin axisymmetric gaseous disk is governed by the Toomre criterion \cite{Toomre1964}:
\begin{equation}
 Q= \frac{c_s \kappa}{\pi G \sigma_g}   < 1 %c_s \kappa/\pi G \sigma_g, 
\end{equation}
where $c_s$ is the sound speed,
$\kappa$ is the epicyclic frequency, and $\sigma_g$ is the gas surface
density. 
This criterion is strictly derived for an axisymmetric, infinitesimally thin disk, and describes the disk's response at a given radius to small (linear) axisymmetric perturbations. 
However, it is often applied to global, vertically stratified disks subject to non-axisymmetric perturbations with considerable degree of success. Once the perturbations grow to nonlinear amplitudes perturbation theory does not apply anymore, and one has to resort to numerical simulations.
Simulations have shown that growing density perturbations lead to collapse only when a second criterion is also satisfied, namely when the cooling time is of order or shorter than the dynamical time within the disk patch satisfying the Toomre threshold \cite{Gammie2001, Rice2003, Boley2007, BoleyDurisen2010, Deng2017}.

Numerical simulations of disks have also shown that protoplanetary disks develop prominent spiral structure even for $Q\lesssim 1.7$, well before $Q=1$ is reached. Spiral structure lead to shock heating of the gas traversing them, and the further growth of the density perturbations requires fast cooling to lower $Q$ to the critical threshold. These non-axisymmetric perturbations can produce disk torques and shocks that
redistribute mass and angular momentum within the disk, and provide a source of heating throughout
gravitationally unstable regions \cite{ZhuHartmann2010}. Thus, spiral arms can act to stabilize the disk by increasing the local sound speed and spreading the disk mass out. In contrast, radiative cooling  decreases the sound speed and destabilize the disk. 

\begin{figure}[h!]
   \centering
    \includegraphics[width=5cm]{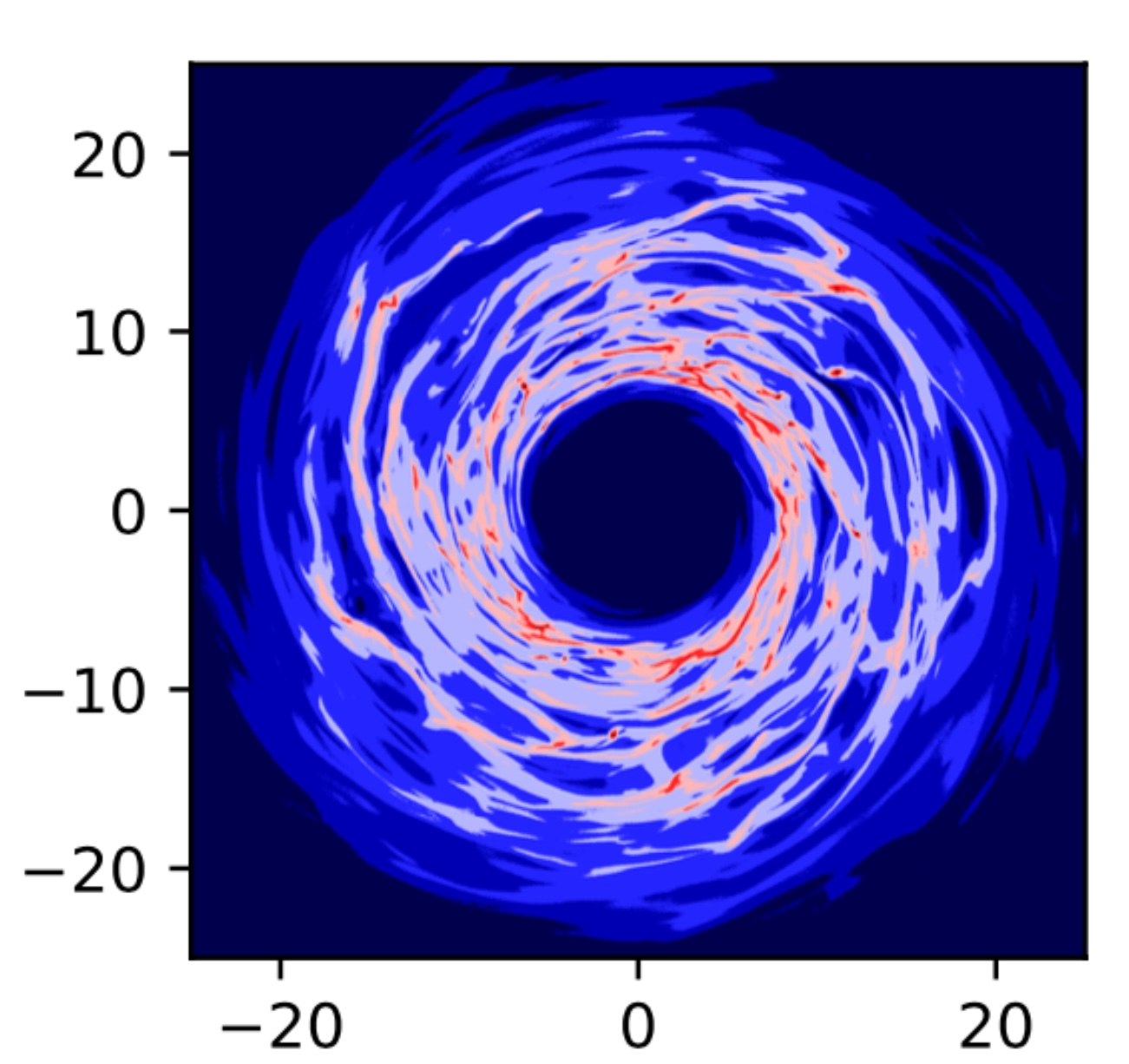}
    \includegraphics[width=6.2cm]{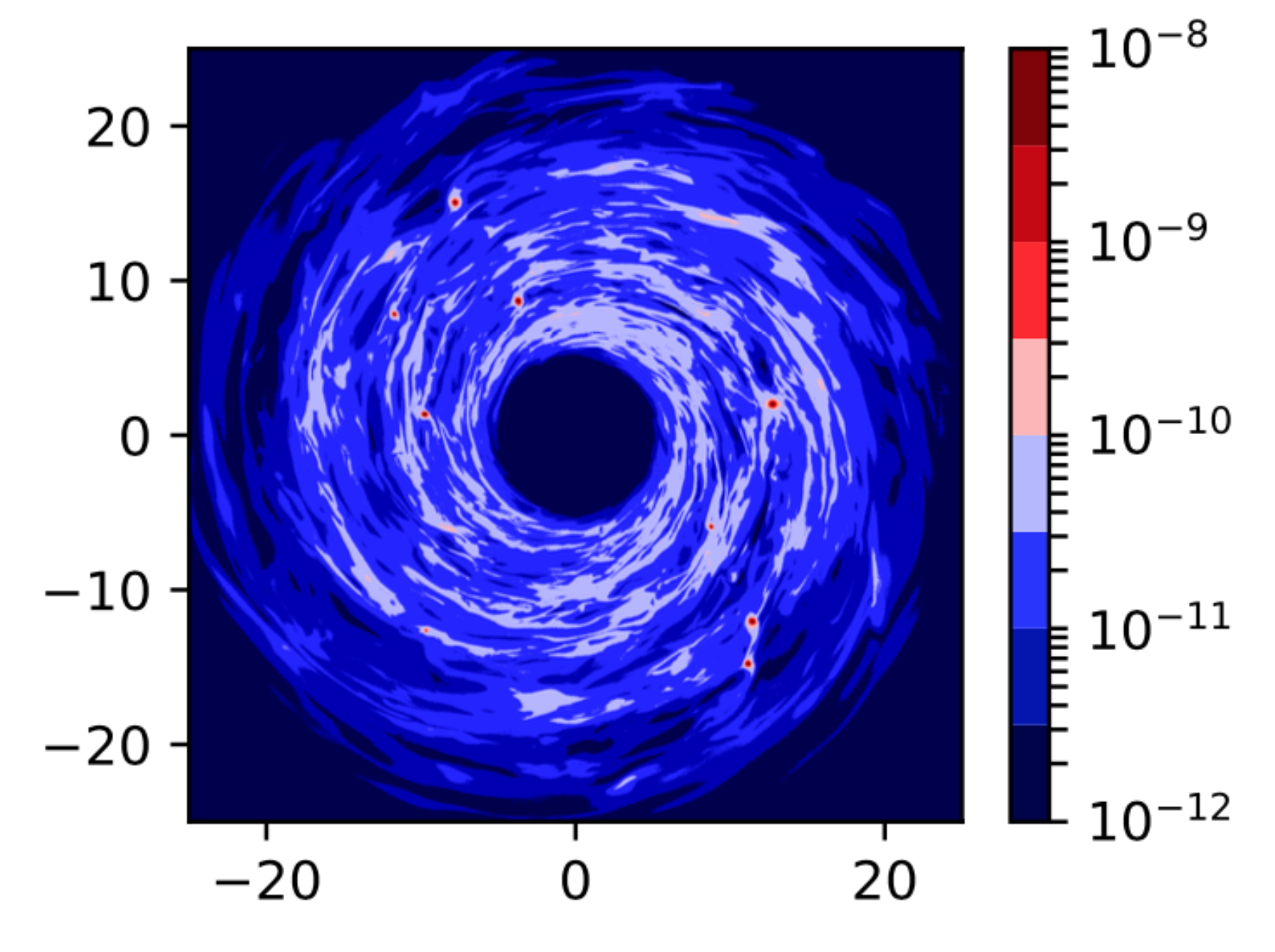}
    \caption[err]{
    {\small
    {\bf Simulations of disk fragmentation.} {\bf Left:} The MHD turbulent disk just before the cooling parameter is switched from 3 to 6.28 at 100 yrs. The simulations correspond to a protoplanetary disk with a mass of 0.07 solar mass (M$_{\odot}$)  spanning 5-25 AU around a solar mass star. Several gravitationally bound clumps with masses of 0.01-0.02 Jupiter-mass are formed as indicated by the midplane density (in g cm$^{-3}$). {\bf Right:} Same as the left panel but for the disk at t=290 yrs. The length unit is 1 AU (adapted from \cite{Deng2021}). }
    \label{fig:deng_mhd_sims}
}
\end{figure}

An equilibrium emerges in the young protoplanetary disk,
where heating and mass transport from the spiral
arms balance cooling and gas infall from the surrounding 
molecular cloud core.
If radiative cooling is efficient, namely if it can dissipate the
viscous heating of the disk on a timescale comparable
or faster than the local orbital time, gravitational instability  leads to a local 
collapse of material in the spiral arms into bound clumps 
\cite{Gammie2001, Deng2017} that can evolve
to become gas giant planets.
Fragmentation can occur both in absence and in presence of 
a magnetic field, this being a relatively novel finding 
shown by recent 3D simulations
of self-gravitating disks which have included a treatment of
both ideal and non-ideal MHD   \cite{Deng2021}.
%When the heating and mass transport from spiral arms can balance cooling and/or infall of gas from the molecular cloud core when the disk is still young,  persistent spiral structure can exist in the disk.  In other cases, cooling and/or mass infall can lead to a secondinstability, i.e., the collapse of regions of spiral arms into bound clumps that can evolve to become gas giant planets. 
Fig.~\ref{fig:deng_mhd_sims} shows an example of MHD simulations of disk fragmentation in which the magnetic field
grows to a dynamically important level owing to a dynamo effect
sustained by gravitoturbulence \cite{Deng2021}. %\commentcm{Comment: this is not a huge deal, but it might be a bit confusing for the reader that we do not speak about MHD in the text and then use an MHD simulation as an example.}
%Fig.~ 4 shows an example of a multi-scale hydro simulation using GASOLINE with a non-ideal gas equation of state to follow a variable mixture of molecular and atomic hydrogen; a clump formed by gravitational instability in a gaseous protoplanetary disk around a 0.3 M$_{\odot}$ star (right) is followed at resolution of better than 10$^{-4}$ AU. A zoom-in of the clump after $\sim$ 1400 years is shown on the left. As can be seen in the left panel, after about 1000 years the clump becomes spherical despite its initial non-spherical shape and about half of its initial mass settles into a circum-planetary disk. 
%It is still unclear under what conditions clumps evolve into gravitationally-bound giant protoplanets. In addition, 
An important aspect is the stellar irradiation of the outer disk.
The nature of this irradiation has not yet been studied in detail and could affect the stability of the disk \cite{Leedham2025}.
The exact conditions that lead to the formation of clumps remain uncertain. In addition, the fragmentation and cooling conditions strongly depend on the disk's properties and are still being investigated. 

\section{Properties of planets arising in DI}
Our understanding of the physical properties of giant planets in the DI model is also incomplete. Constraining these properties, however, will allow one to perform meaningful comparisons with observations and helps to distinguish DI from other formation pathways. In addition, inferring the expected masses, compositions, and orbital configurations of such planets can lead to insights into the physical conditions under which disk fragmentation occurs and how these conditions influence the resulting planetary population. 

\subsection{Masses}
The initial masses are expected to be on the order of a few Jupiter masses, which is  close to the Toomre mass\footnote {The Toomre mass is the mass associated with the most unstable scale in a disk that becomes gravitationally unstable. It is typically derived from the Toomre $Q$ parameter, hence it is computed in the simple framework of linear perturbation theoyr. A simplified expression for the Toomre mass is: $M_T \sim \Sigma \lambda_T^2$, where $\Sigma$ is the disk's density and $\lambda_T$ is most unstable wavelength from the Toomre criterion. Typical Toomre masses in disks are of the order of a Jupiter mass.} for typical
disk parameters. Simulations have shown, however, that clumps form
in overdense regions inside spiral arms, not in the axisymmetric
disks as assumed in the Tooomre instability theory \cite{Durisen2007}. Shock dissipation in spiral arms reduces the characteristic
scale of fragmentation by factors of a few, allowing Jupiter-sized and
even smaller clumps to form \cite{BoleyDurisen2010}.
Early simulations suggested that, after a few disk orbits following fragmentation, clump masses range between about 1 and 10 Jupiter masses (\SI{}{\MJ}), mainly as a result of growth via gas accretion.
However, the inferred masses strongly depend on the assumed conditions within the disk, on gas thermodynamics, and even on the effect
of magnetic fields, which affect
gas accretion onto the clumps, and thus their further growth.
Currently, the  expected initial mass of clumps is unknown.  Different studies predict different masses that can depart significantly from the Toomre mass (see e.g., \cite{2010Icar..207..509B,Forgan2011,Schib2021} and references therein). 
The value from \cite{2010Icar..207..509B}, which is typically of the order of a Jupiter-mass, is in good agreement with their radiation hydrodynamic simulations as well as with \cite{Tamburello2015}.
On the other hand, global simulations of gravitational instability with full radiation transport suggest that the distribution of initial fragment masses is centered at \SI{20}{\MJ} \cite{Xu2025}. However, the latter simulations do not include adaptive resolution, which may lead to the formation of smaller fragments \cite{Durisen2007}.
Since
most 3D simulations published in nearly two decades do not run for 
enough time to approach a stage at which gas accretion saturates, it is not yet possible to predict the final masses of the protoplanets. In addition, clumps can also lose mass due to various effects described below,  which can be severe if they migrate rapidly inward. This hinders even more a robust prediction of the final masses.

%Several mechanisms contribute to mass accretion and mass loss during the formation of clumps. 
The most significant mechanism of mass growth is gas accretion, whereby clumps attract gas from the surrounding disk due to their gravitational pull, leading to their continued growth \cite{ZhuHartmann2012,Schib2022}. In some cases, clump mergers also plays a key role, where multiple fragments merge into a more massive object \cite{Mayer2002}. Furthermore, solid accretion can also increase the clump's mass, and also change their bulk compositions (e.g., \cite{Helled2014,BaehrKlahr2019ApJ}). 

Clumps may experience mass loss due to interactions with other objects in the disk \cite{Boss1997}. Additionally, disk photoevaporation and outflows can also lead to envelope stripping, resulting in further mass loss. 
Finally, tidal downsizing could also play an important role.  Tidal interactions can lead to the stripping of material from the clump, leading to the formation of less massive objects, such as planetary cores or lower-mass gas giants \cite{Nayakshin2010}.

Tidal downsizing is also a potential explanation for the diversity in planetary system architectures. In this scenario, only the clumps that form in more stable regions of the disk or those that undergo fewer tidal interactions may grow large enough to become gas giants, while others may remain much smaller in mass or fall into their host stars.

\subsection{Clump Evolution}

The evolution of proto-planetary clumps formed in DI can be divided into three main stages.  
During the first stage, also known as ``pre-collapse", forming companions are cold and extended clumps, having low densities and sizes of a few thousand to ten thousand times Jupiter's present radius (\SI{}{\RJ}).
Shortly after their formation, the internal pressures and temperatures within the clumps are sufficiently low that hydrogen is in molecular form. 
During the pre-collapse stage, clumps are vulnerable to destruction due to tidal disruption and disk interactions. 
As the clumps radiate energy and contract, their internal temperatures increase.
Once the central temperature reaches $\sim \SI{2000}{\kelvin}$, the molecular H dissociates, and a dynamical collapse of the entire object occurs.
This second phase of collapse due to H disassociation is known as ``dynamical collapse". The third phase, known as post-collapse, corresponds to the long-term cooling and contractions of the planet which takes billions of years.
During this stage, the companion is compact and tightly gravitationally bound and resembles a giant planet that formed by the core accretion model. 
 \par 

\begin{figure}[h!]
   \centering
    \includegraphics[width=11.5cm]{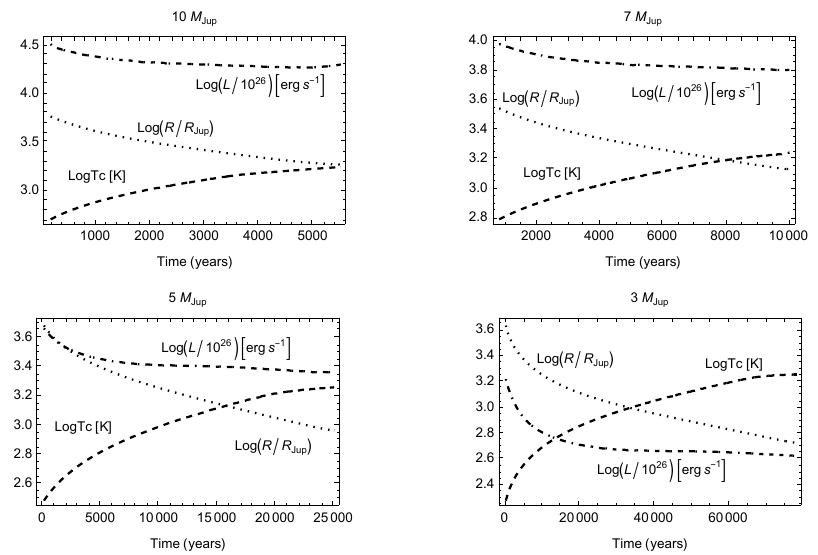}
    \caption[err]{
    {\small
    Pre-collapse evolution of \SI{3},{5}, {7} and \SI{10}{\MJ} evolving in isolation, assuming solar composition and interstellar dust opacity. The luminosity, radius, and central temperature vs.~time are shown. Adapted from \cite{helled_metallicity_2010}.
    }
    \label{fig:clump_evolution_helled2010}
}
\end{figure}

More massive clumps are larger in size and contract on shorter timescales.
They also have higher internal temperatures.
As a result, more massive clumps are less susceptible to destruction during their early evolution.
The pre-collapse evolution of clumps with masses of \SI{3}, \SI{5}, \SI{7} and \SI{10}{\MJ} is presented in Fig.~\ref{fig:clump_evolution_helled2010}. We note, that clumps with different composition are expected to have different evolution history and contraction timescales \cite{Humphries2019}. In addition, the timescales of the different phases strongly depend on the clumps' initial masses and the local conditions in the disk, as well as the physical processes at play (migration, accretion, mass loss, etc). The pre-collapse timescale can vary between a few times
$10^3$ to millions of years depending on the planetary mass, metallicity, and distance from the star, as well as clump-clump interactions.

The initial composition of the clumps is often assumed to be similar to stellar, although it was shown that clumps could be enriched with heavy elements from birth due to the concentration of dust around spiral arms, where fragmentation occurs  \cite{BoleyDurisen2010,Boley2011, BaehrKlahr2019ApJ} or due to a different concentration of chemical species that are abundant in  spiral shocks within the disc \cite{Ilee2017}. 
In addition, during the pre-collapse phase, the composition and internal structure of the clumps can change due to grain settling and planetesimal/pebble accretion. These processes can lead to a significant change in the bulk composition of the clump and also lead to the formation of heavy-element cores in the deep interiors (see \cite{Helled2014,Nayakshin2014} and references therein.
The pre-collapse timescales depend on the physical properties of the clumps. For example, the pre-collapse timescale can be reduced if the protoplanets have
low opacity (and therefore more efficient cooling), due to either low metallicities or due to opacity reduction via grain growth and settling \cite{Helled2011,Mordasini2014}. The pre-collapse timescale can also be shortened due to collisions between clumps \cite{Matzkevich2024} (see Sect.~\ref{sec:ccoll} for details). On the other hand, higher opacity leads to slower cooling and contraction therefore increasing the pre-collapse timescales. It is therefore clear that in order to predict the evolution of clumps (and giant planets in general), a good understanding of their compositions  and atmospheric conditions is required. 

In the following sections we discuss recent developments in modeling disk instability conducted by various groups associated with PlanetS. 

\subsection{Magneto-hydrodynamic simulations}
Magnetic fields influence angular momentum transport, turbulence, and disk stability - factors thar are critical to assess the conditions for disk fragmentation. As a result, magneto-hydrodynamic (MHD) simulations are needed to capture these effects, which pure hydrodynamic models cannot. The spiral waves sustained by gravitational instability drive a magnetic dynamo (different from the 
MRI) \cite{RiolsLatter2018Magnet, RiolsLatter2019},
shown by 3D-simulations \cite{Deng2020}.
The resulting magnetic fields are resilient to resistivity and are strong, 
reaching $\beta_\text{plasma} \approx 10$.
This mechanism also works in weakly-ionized disks (contrary to MRI)
for values consistent with ionization levels measured in protoplanetary
disks \cite{RiolsXu2021}.
Such an ionization may arise from the central star,
other close stars, cosmic rays or even spiral shock heating
\cite{Podolak2011}. Additionally, 3D simulations of protostellar collapse
from realistic magnetized turbulent molecular clouds find that
protostellar disks, already at birth, are threaded
by a magnetic field with a strength similar to what is found as
a result of the spiral-driven dynamo. 

\par 

Recently, \cite{Deng2021} and \cite{Kubli2023} investigated clump formation using magneto-hydrodynamic simulations of protoplanetary disks incorporating a spiral-driven dynamo.
Their disks had a mass of $m_\text{disk} = 0.1M_\text{sun}$ around a solar-mass star
with cooling modeled using beta-cooling \cite{Gammie2001, Deng2017} which induces gravitational instabilities and disk fragmentation. They found that fragmentation leads to the formation of long-lived, bound protoplanets with masses significantly smaller than those predicted by conventional disk instability models. The initial mass of the fragments is significantly smaller than that predicted via the conventional Toomre mass. This is because the magnetic field affects global spiral structure in the disk, shifting dominant modes to smaller wavelengths, and locally, in regions of lower shear, where collapse is most likely, it reduces the Coriolis force,
promoting collapse on small scales that would otherwise be stable \cite{Kubli2023}.

These lighter clumps survive shear and do not grow further. This can be attributed to a shielding effect of the magnetic field. The magnetic field strength thereby accumulates around the clumps, probably due
to its rotation. This inhibits the local inflow of matter. In the clump's interiors, however, the gas pressure
is the dominating stabilizing force.

The study suggests that intermediate-mass, gas-rich planets are more common, while gas giants are rarer, aligning with the observed mass distribution of exoplanets. 
This research provides new insights into planet formation, highlighting the role of magnetic fields in producing intermediate-mass planets and offering a potential explanation for the observed distribution of exoplanet masses. Future research should include various disk properties and account for the heavy elements that do not only influence the disk and clump evolution, but also the final properties of the clumps. 

\begin{figure}
\centering
\includegraphics[width=0.8\linewidth]{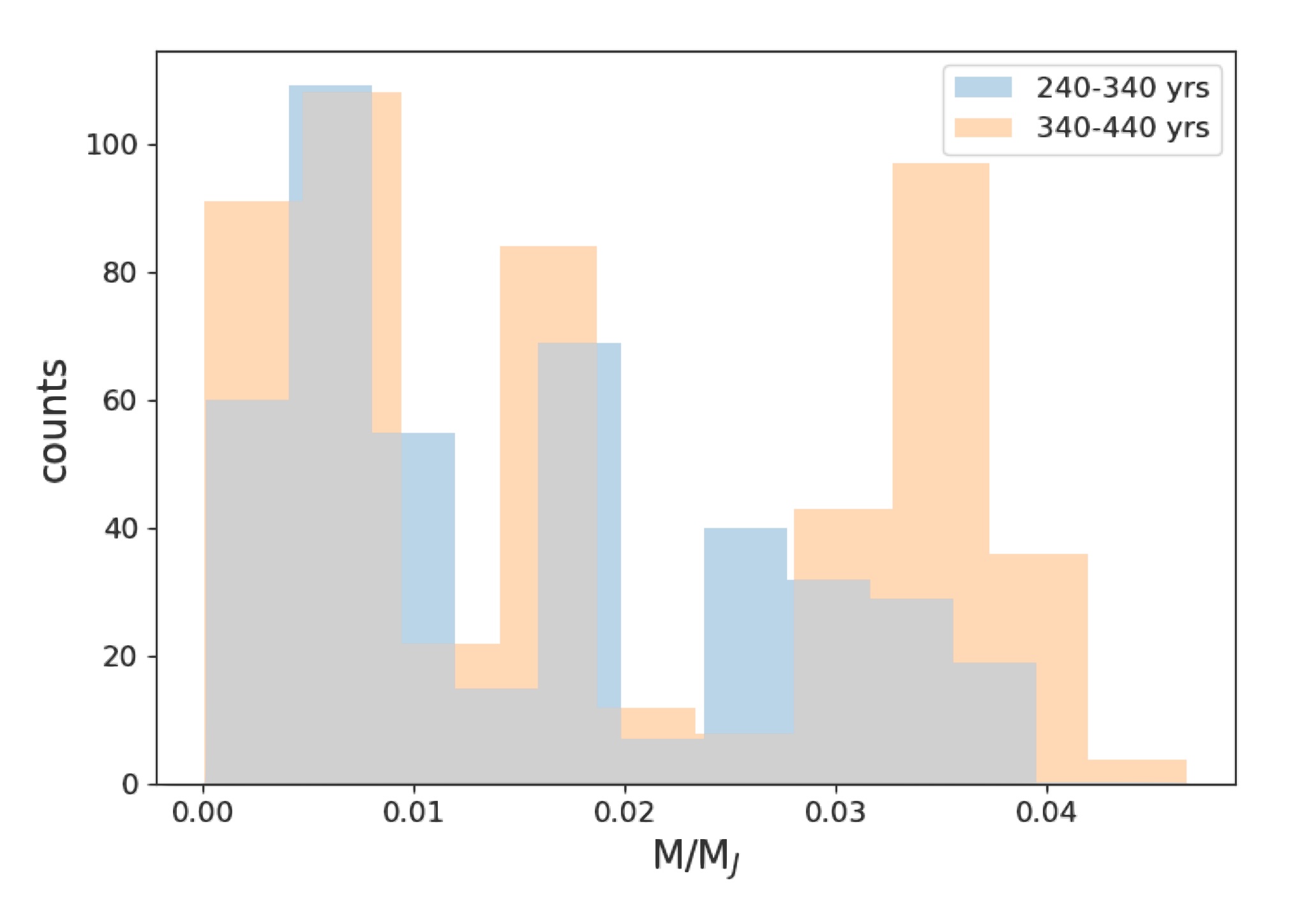}
\caption{\label{fig:deng_mass_frequency}The inferred frequency of planets of different masses from \cite{Deng2021}. The planets are identified in 60 consecutive snapshots taken between 240-340 yr and 340-440 yr of the main MHD simulation. The width of the histogram bins is \SI{0.004}{\MJ}}. 
\end{figure}

\begin{figure}
\centering
\includegraphics[width=0.49\linewidth]{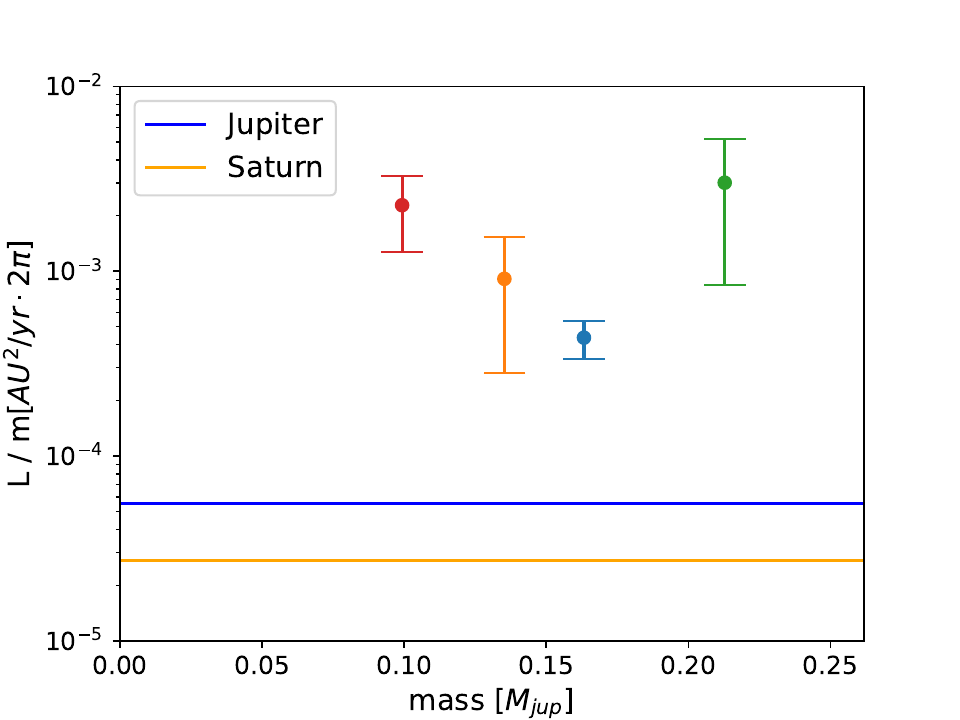}
\includegraphics[width=0.49\linewidth]{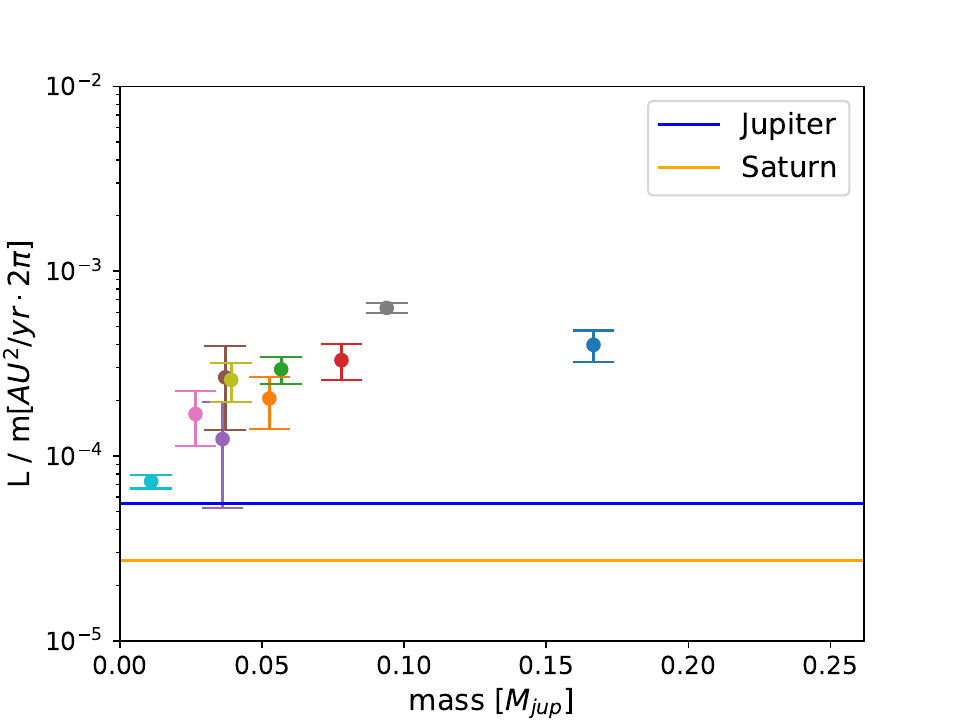}
\caption{
The specific angular momenta 
of the clumps in the magnetized
case (right) are significantly lower than
in companion non-magnetized
simulations (left);
thereby being closer to Jupiter's
and Saturn's.
Figure taken from \cite{Kubli2023}, licensed under CC-BY 4.0.
}
\label{fig:angmom}
\end{figure}

The clumps exhibit rotation leading to an 
oblate shape.
Comparing the MHD simulations
to HD simulations (that neglect the magnetic field) 
also reveals that clumps have significantly less
angular momentum 
in the magnetized case,
which is shown in Fig.~\ref{fig:angmom}.
It has long been the case 
that 
the angular momentum of clumps
originating from GI is much higher
than the observed angular momenta in our solar system
\cite{Durisen2007}.
In the simulations including
the magnetic fields, the 
angular momenta
match
much more closely the spin of giant planets on our Solar System as
well as the constraints on the latter in extrasolar giants.
The reason for this difference
could come from resistivity
leading to a dissipation of kinetic energy
via the action of the magnetic field.

\begin{figure}
\centering
\includegraphics[width=0.76\textwidth]{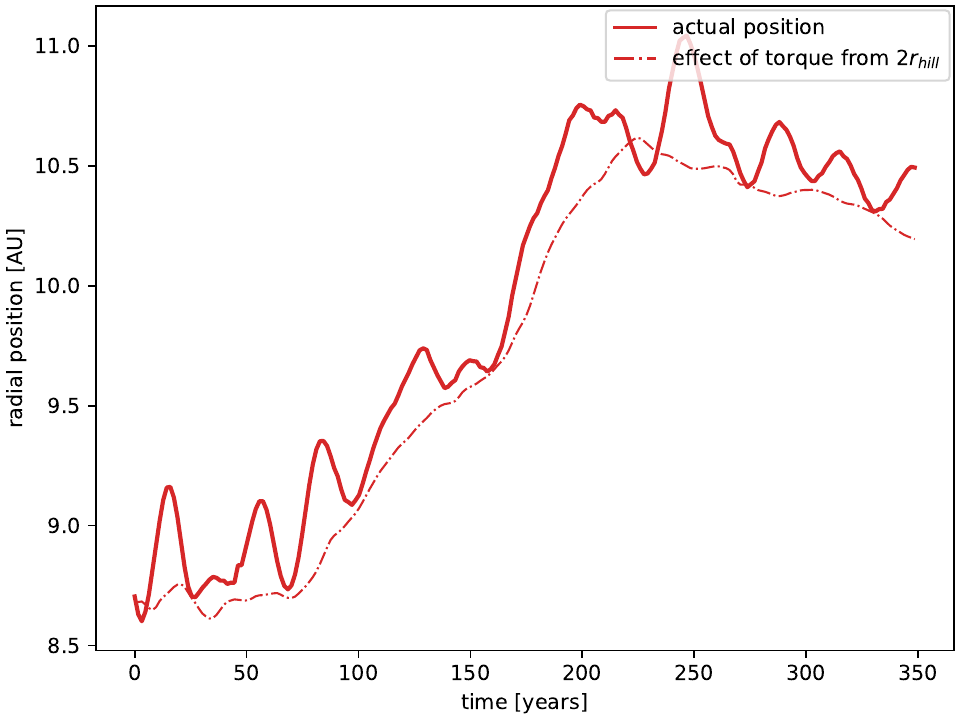}
\caption{Evolution of the radial position of a clump over time.
The torque arising from a region of $2r_\text{hill}$ around the clump's orbit
seems to account for most of the migration besides clump-clump interactions.}
\label{fig:migration}
\end{figure}

In \cite{Kubli2025}, their migration was investigated, 
one aspect of the longer-term evolution
of such clumps.
They traced their position and computed the 
gravitational torque arising
from different components of the disk.
The clumps in the simulations were not replaced by sink particles,
but were the original clumps arising from disk instability;
therefore they can interact with the surrounding flow and 
evolve internally.
The radial positions of the clumps changes
fast, on an orbital timescale, and 
can be directed both inward or outward.
Migration has different causes in such disks:
Direct interactions between two clumps and interactions
with the disk's structures.
Clump-clump interactions lead to very fast changes in radial position,
however over the long-term are less important than torques \cite{Schib2022} arising from the disk's material.
The disk flow is highly turbulent and exhibits a spiral pattern,
modified by the magnetic field.
The resulting migration is stochastic but shows 
persistent features
over multiple orbits.
An example of the evolution of the radial position of a clump is 
presented in Fig.~\ref{fig:migration}. For this clump no encounters
with other clumps with other clumps occur.
Kubli et al. (2025) \cite{Kubli2025} showed that most of the effect of migration,
in absence of clump-clump interactions,
could be explained by a gravitational torque arising from 
$\approx 2$ Hill radii apart from the respective clump's orbit.
Additionally, migration can be both inward and outward, as shown
in the Figure.
These results suggest that the planets resulting from GI
would be distributed over a wide range of radii rather than
near their birthplace. This implies that wide-orbit planets 
should only be a subset of the population of planets formed
by DI, and that planets at small radii, a few AU or less,
including hot analogs of hot Jupiters, can also arise, as originally argued using semi-amalytical models in Galvagni \& Mayer \cite{GalvagniMayer2013}
and Müller et al.\cite{Muller2018}. In addition, Kubli et al. \cite{Kubli2025} did not find any ejection of clumps, 
a phenomenon
previously observed in simulations employing point-like sink particles \cite{Boss2023}
that did not resolve the clumps' internal structure and hence
the effect of tides on extended objects.
Such ejections, if they occur, could contribute to the population
of free-floating planets 
(observed e.g. in \cite{MiretRoig2022}).
These results show, however, that such ejections are unlikely in the
early phases of disk instability
captured in the simulations.
As mentioned above, after the dynamical collapse the protoplanets have much smaller radii
and undergo stronger gravitational scattering during encounters,
which could result in ejections, especially for wide-orbit protoplanets beyond 50 AU from the star where  the binding energies are significantly lower. This second stage is too advanced in the time evolution and requires too high resolution to be included self-consistently in 3D hydro or MHD simulations. These later stages must be modeled differently, typically using 1D evolution models and population synthesis modes as we discuss below.  

\section{Population Synthesis models}
The key challenge in the DI model is to understand how these clumps evolve thermally and dynamically, and how their growth can lead to the formation of companions that may match some properties of the observed population. Population synthesis models have been developed to study this problem by simulating the evolution of a large number of systems, using simplified representations of physical processes that govern formation.

The basic idea behind population synthesis in the context of DI is to create a statistical ensemble of star-and-disk systems with varying initial conditions, such as disk mass, temperature, and metallicity, and to track the formation and evolution of clumps in these disks. These models use a set of simplified assumptions about the physics of gravitational instability, thermal evolution, and accretion to simulate how clumps form, evolve, and potentially become giant planets. The models typically solve the equations of motion for a set of clumps, accounting for gravitational interactions, hydrodynamical processes, and radiative cooling. By running these simulations with a variety of parameters, population synthesis models provide statistical predictions for the mass, orbital parameters, and occurrence rates of companions. These predictions can then be compared to observational data to assess the viability of the instability model as a mechanism for planet formation.

The population synthesis method is well established in the core accretion paradigm (see e.g. \cite{IdaLin2004,Mordasini2009,BurnMordasini2024}).
 For DI, some models exist, but they are less mature. 
Forgan \& Rice \cite{Forgan2013,Forgan2018} 
investigated DI in pre-evolved disks around sun-like stars.
Clumps with a given mass \cite{Forgan2011} were added to the system at random locations and allowed to evolve, including orbital migration.
They also considered core formation and mass-loss and found that the survival of a close-in terrestrial planet is a possible, but very unlikely outcome. Gas accretion was not included, but the second publication includes gravitational interaction between fragments. The resulting population was found to be dominated by massive giant planets and brown dwarfs at wide separations.

Another population synthesis project was performed by S.~Nayakshin \cite{Nayakshin2015}.
It studied the evolution of fragments formed by DI at random locations in the disk, though here the disk's evolution was also considered.
Only one fragment per system was simulated, with accretion of pebbles, orbital migration, gap formation and the possibility of tidal downsizing.
The surviving population also contained many massive giants on wide orbits.
However, here a substantial number of terrestrial planets resulted from tidal downsizing and survived close to the host star, in contrast to what was found by \cite{Forgan2018}. Currently, it seems that terrestrial planet formation is not a likely outcome of DI. Overall, the mentioned population synthesis projects represent important steps in the direction of understanding the expected  population in the DI model. 

Recently, Schib et al.  \cite{Schib2025a,Schib2025b} investigated the fragmentation of circumstellar disks and clump evolution in great detail \cite{Schib2021, Schib2022, Schib2023}. They studied initial conditions for the formation of companions in DI in a statistical sense with the aim to make quantitative predictions that can be tested observationally.
It was confirmed that fragmentation occurs when disks are young and massive.
Therefore, they also included the formation phase of the star-and disk system by considering the infalling mass from the molecular cloud core. By applying results from hydrodynamical simulations \cite{Hennebelle2016, Bate2018}, 
\cite{Schib2021} investigated the condition for disk fragmentation in a large parameter space in final primary mass, from the brown dwarf to the B-star regime.
It was found that the infall radius, i.e. the location in the disk where the infalling material hits the disk is a crucial parameter. 

They also showed that infall radii motivated by simulations of non-magnetized cloud collapse are larger, whereas those  motivated by MHD-driven collapse are too small, when compared to observed Class~0 disk sizes \cite{Tobin2020}.
The former were found to fragment excessively, the latter not at all.
Furthermore, the importance of gas accretion and orbital migration for clump  survival was realized.
In a dedicated study \cite{Schib2022}, a model for self-consistent planet-disk interaction was developed and tested against 3D hydrodynamical simulations \cite{DAngelo2008,DAngelo2010} including exchange of mass and angular momentum.
An additional study \cite{Schib2023} was conducted to address the question of the early disk size.
A population synthesis of circumstellar disks was performed that exhibits a distribution of early disk sizes compatible with observed Class~0 disks, while at the same time producing a distribution of final primary masses according to the observationally accessible initial mass function \cite{Chabrier2005}.
It was shown that in addition to the early disk size, the efficiency of accretion heating also plays a key role in disk fragmentation.
When chosen in such a way that the total luminosity of young disks are consistent with  observed luminosities \cite{Tobin2020}, it was found that about 10\% of the systems fragment.

\begin{figure}
\centering
\includegraphics[width=1.\linewidth]{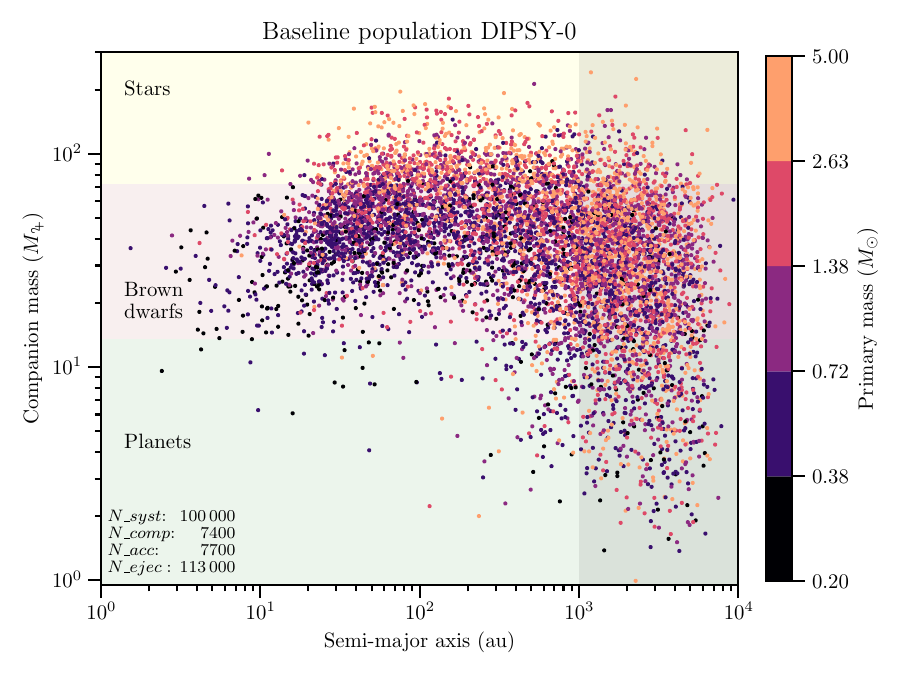}
\caption{\label{fig:diam}Mass vs. semi-major axis of surviving companions in the baseline population DIPSY-0 \cite{Schib2025b}.
The colors give the final mass of the primary (host star).
In the bottom left corner, the total number of systems, the number of surviving, accreted and ejected companions is given.
The figure shows all surviving companions without taking into account the different relative abundances of primaries with different masses according to the stellar initial mass function (from \cite{Schib2022}, licensed under CC-BY 4.0). }
\end{figure}

These models were then combined into an extensive DI population synthesis model (named ``DIPSY"), as described below. We note that the work by O.~Schib and the DIPSY project is the result of a new initiative supported by PlanetS, led by C.~Mordasini \& R.~Helled.
The model includes the formation of the star-and-disk system and the evolution of the disk until its dispersal. 
When a disk is found to fragment, one or several clumps are now inserted into the disk.
These clumps can interact with the disk following \cite{Schib2022} and the clumps' internal evolution is followed by interpolating pre-calculated tables of isolated clumps (\cite{Humphries2019}) that includes the dynamical collapse. 
The clumps' evolution is influenced by their surrounding disk through radiation \cite{Vazan2012, Humphries2019} and they can lose mass to the disk if their radius exceeds their Hill sphere.
The three-dimensional gravitational interaction of the forming companions was followed by using an n-body integrator \cite{Chambers1999}.
Such interactions can lead to the ejection of companions or to their collision (see also Sect.~\ref{sec:ccoll}).
Companions can also accrete on the primary by inward scattering or gas disk-driven orbital migration. 

The results of \cite{Schib2025a,Schib2025b} are based on the initial conditions and the physical mechanisms described above. The first paper describes in detail the model, including tests and example applications. 
The second paper describes the results of a population synthesis calculation for 100,000 systems as well as several additional populations. 
A key result of these new population models is presented in Fig.~\ref{fig:diam}. In this figure, all the surviving companions of the nominal DIPSY-0  population are shown, with the primary mass color coded.
Note that the different relative abundance of primaries of different masses is not considered.  
The figure suggests that DI forms companions from the stellar mass down to the planetary mass regime, dominated by massive brown dwarfs.
Planetary mass companions seem to be rare, and typically at wide separations due to scattering by more massive companions. 
Survivors are found to have on average masses $\sim$ \num{25} times their initial clump mass. This demonstrates the key importance of two processes: gas accretion and clump mergers are crucial in shaping this population.
These effects must be studied further to assess their impact on the inferred population (see next section). This demonstrates the important interplay between detailed (often 2D/3D) models and complementary global planetary population synthesis models (often low dimensional). 

\section{Clump collisions}\label{sec:ccoll}

Clump-clump interactions play a significant role in determining the final population of planets formed by DI. So far, most studies either neglect clump-clump interactions or treat the clumps as point masses (e.g., \cite{Forgan2013, Forgan2018, Muller2018}) and assume that collisions lead to perfect merging. However, it was recently shown by \cite{Matzkevich2024} that these simplified assumptions are inappropriate for modeling clump collisions, especially during the pre-collapse stage when the clumps are very extended and only weakly gravitationally bound.  
\par 
The study finds that the outcomes of clump collisions are extremely diverse (see Tab.~\ref{tab:def_collision_outcomes} for a definition of the outcomes and Fig.~\ref{fig:collision_outcomes_matzkevich2024} for examples): they range from perfect merging, erosion and disruption to hit-and-run collisions. For the parameter space investigated in this study, perfect mergers were found to be  rare. Mergers typically occurred in encounters below the mutual escape velocity and at intermediate impact parameters. Collisions typically lead to mass loss and erosive and disruptive outcomes are very common. Due to the shallow gravitational potential of the clumps, disruptive events (where more than 50\% of the total mass is lost) can already occur for impacts with rather low relative velocities of a few \SI{}{\kilo\meter\per\second}. 

\begin{figure}
\centering
\includegraphics[width=0.8\linewidth]{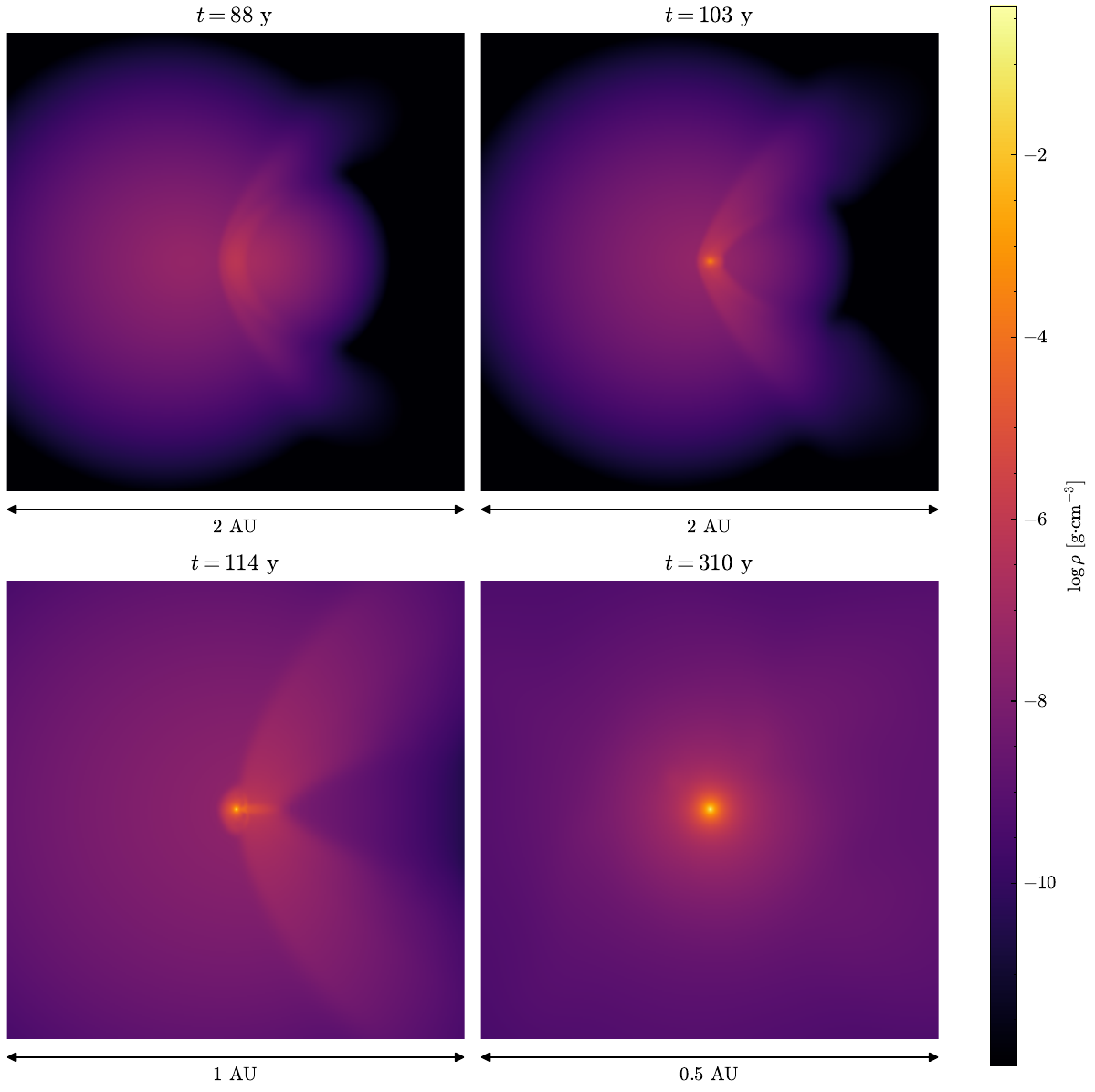}
\label{fig:collision_collapse_matzkevich2024}
\caption{Snapshots showing a head-on collisions between evolved 10 and \SI{5}{\MJ} clumps with $v_{imp}/v_{esc} = 1.05$ that initiates dynamic collapse. During the impact a compact region behind the shock front is formed where the density increases by several orders of magnitude (top panels).  Later the collapsed region pierces through the shock front (bottom left panel) and regains hydrostatic equilibrium (bottom right panel). After the collision \SI{77}{\percent} of the total colliding mass remains bound. Such collisions are a pathway to substantially shorten the pre-collapse stage of a clump and increase their survival probability. Figure from \cite{Matzkevich2024}, licensed under CC-BY 4.0.}
\end{figure}

\begin{table}
\label{tab:def_collision_outcomes}
\centering
\begin{tabular}{lccc}
\\
\hline\hline
Description & $N_{\rm frag}$ & $M_{\rm bound}/M_{\rm tot}$\\
\hline
Perfect merger & 1 & $\geqslant$ 0.95 \\
Erosion & 1 &  $\geqslant$ 0.5 \\
Disruption & 0/1 & $<$ 0.5 \\ 
\hline
Perfect hit-and-run & 2 &  $\geqslant$ 0.95 \\
Erosive hit-and-run & 2 & $<$ 0.95\\
\hline
\hline
\\
\end{tabular}
\caption{Classification of the collisions depending on the final number of bound fragments as well as ratio of final total bound mass to initial total mass. By definition, a perfect merger is a collision where 100\% of the colliding mass remains bound in a single fragment. Here we consider a perfect merging as a collision with a mass loss of up to 5\% (from \cite{Matzkevich2024}, licensed under CC-BY 4.0).}
\end{table}

As a result, collisions resulting in substantial mass loss are expected to be common even at large distances from the central star. If the impact angles are large and the impact velocities exceed the mutual escape velocity, hit-and-run encounters occur. In these cases, both bodies survive the collision but can exchange mass and angular momentum. Furthermore, during the passing of the shock, the core temperature of a clump can be temporarily very high and reach values above the hydrogen dissociation limit. As a result, collisions can initiate dynamic collapse and shorten the time required to reach dynamic collapse over a wide range of impact conditions (see Fig. \ref{fig:collision_collapse_matzkevich2024} for an example). Accelerated dynamic collapse due to impacts increases the survival probability of clumps since during the pre-collapse phase they are most susceptible to disruption. This in turn could have a substantial impact on the inferred population of giant planets in the disk instability model. 

It is yet to be determined whether clump mergers are a common outcome. Mergers could be more common if relative velocities are sufficiently low  (as found by Schib et al.~2025). Future studies are required to determine the typical impact conditions of clumps in protoplanetary disks. 
Dedicated SPH simulations can then be performed to investigate the collision outcomes under such conditions.
We suggest that future population synthesis studies include different collision outcomes depending on impact parameter and relative velocity based on the SPH results.

\begin{figure}
\centering
\includegraphics[width=0.48\linewidth]{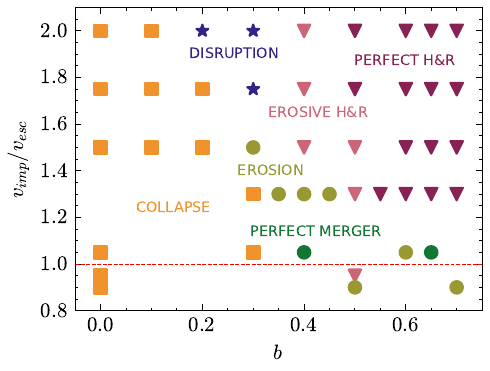}
\includegraphics[width=0.48\linewidth]{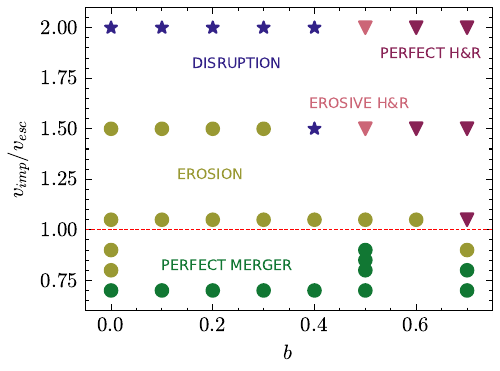}
\label{fig:collision_outcomes_matzkevich2024}
\caption{The diverse outcomes of collisions between “evolved” (but still before dynamic collapse) 10~M$_J$ (left) and “young” (after becoming gravitationally bound) 3~M$_J$ (right) clumps inferred from 3D impact simulations. Perfect merging is found to be a rather rare outcome and collisions often result in erosion / disruption of the clumps or hit-and-run encounters (where two clumps remain, exchanging some mass and angular momentum). Collisions can also shorten the pre-collapse stage during which clumps are most likely to be disrupted (see Fig.~\ref{fig:collision_collapse_matzkevich2024} for an example). Figure from \cite{Matzkevich2024}, licensed under CC-BY 4.0.}
\end{figure}

\section{Observational Constraints}
Understanding the connection between theoretical models of planet formation and observational data is essential for evaluating the viability of the DI model. This involves comparing predictions regarding the formation timescales, occurrence rate, mass distribution, and orbital characteristics of giant planets and other sub-stellar companions with the observed properties of protoplanetary disks and exoplanetary systems. By identifying key observational signatures that are consistent with rapid gravitational fragmentation in massive, cold disks, we could assess the extent to which DI contributes to planet formation across different stellar environments.

Observations of both disks and protoplanets provide the potential to test and refine the disk instability models. In the past decade, our understanding of protoplanetary disks advanced substantially, in particular thanks to the Atacama Large Millimeter/submillimeter Array (ALMA) and high-contrast imagers like VLT/SPHERE. Recently, the James Webb Space Telescope started providing data that will enable to test theoretical predictions at larger separations while the innermost regions of planetary systems can  be investigated by upcoming ELT instruments like HARMONI and METIS (see Direct Imaging chapter from Cugno \& Meyer).

\subsection{Protoplanetary disks}
The presence of substructures in protoplanetary disks is now considered ubiquitous. These structures are seen in the mm dust continuum with ALMA, in optical and NIR high-contrast imaging observations, and more recently as gas kinematic perturbations in the disk gas velocity field (e.g., \cite{Andrews2018, Oberg2021, Pinte2023}). Most of these structures in the outer regions of the  disks have been successfully modeled by hydrodynamic simulations and explained (among other possibilities) by protoplanets, which could have been formed by DI. 

Many rings and gaps align well with predictions from the core accretion model, where growing planetary cores carve gaps in the disk. However, some observed gaps are located at large orbital distances ($>50$~AU), where core accretion is expected to be inefficient due to long formation timescales, potentially favoring the disk instability model. One example is the AS209 disk, where a large ($\sim80$~AU in width) gap was detected at a separation of 210 AU \cite{Guzman2018}, and a candidate circumplanetary disk was found in the middle of it \cite{Bae2022}. Unfortunately, direct imaging efforts to date have failed to confirm this candidate \cite{Cugno2023_ISPY, Cugno2023_magao}.

\begin{figure}
    \centering
    \includegraphics[width=0.9\linewidth]{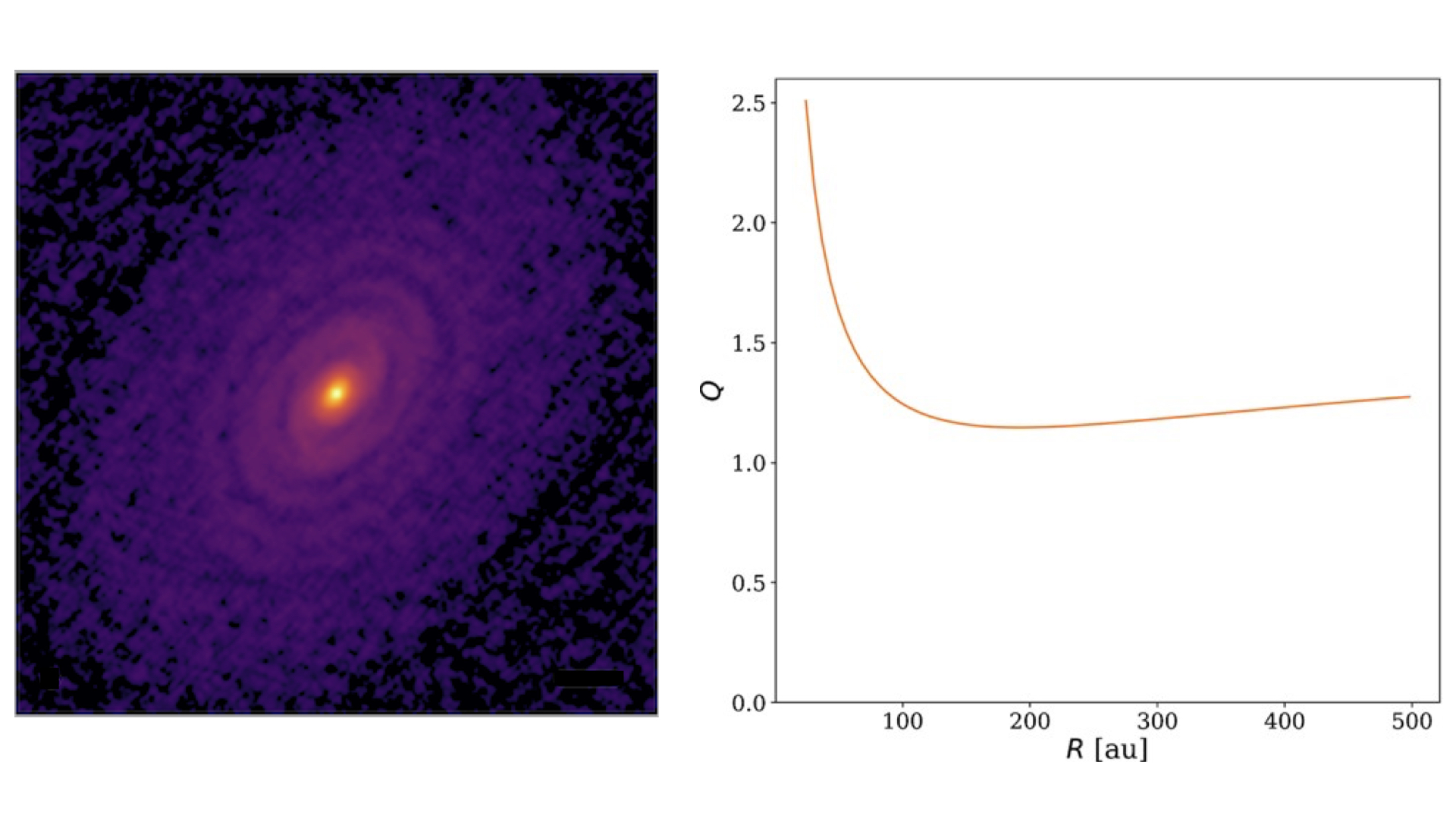}
    \caption{{\it Left:} Dust continuum image of the IM Lup disk, observed in Band 6 by the DSHARP project \cite{Andrews2018}.  {\it Right:} Toomre $Q$ parameter vs.~separation for the IM Lup disk, as inferred from dynamical mass measurement of the disk \cite{Lodato2023}.}
    \label{fig:disk_mass}
\end{figure}

In addition, the presence of spiral arms is one of the most compelling features that are often interpreted as the signature of DI. As suggested by Equation 1, disk fragmentation can occur when the disk's density is sufficiently high, i.e., when the disk is massive. Unfortunately, the disk mass has long been debated and until recently, there was an ambiguity in the measurements because they relied on model assumptions. Currently,  thanks to high-resolution observations, it is possible to infer a dynamical measurement of the disk mass. This is done by modeling the disk's rotation curve, and accounting for both the gravitational influence of the central star and the self-gravity of the disk itself (e.g., \cite{Veronesi2021}). More accurate disk mass measurements can be used to identify the best candidates for disk fragmentation and assess whether spirals are indeed associated with a gravitational collapse.\\

Spiral arms can also arise in the context of planet-disk interactions, where a forming object carves out spiral density waves in the disk. A typical example is the disk around HD100453, which shows two prominent spirals driven by a low-mass stellar companion. Unfortunately, for many other spiral disks, no clear companion has been detected so far (e.g., \cite{Cugno2024a}). Only one candidate protoplanet associated with the spiral arm in MWC758 has been proposed through LBT/LMIRCam observations \cite{Wagner2023}. 
It is difficult to distinguish between spirals caused by planets and those caused by a gravitational instability, but several diagnostics have been proposed. One approach is to measure the rotational velocity of the spirals: if driven by one or more companions, the spirals should move at the keplerian velocity of the driver, while if created by disk instability they should move at the expected keplerian velocity at the separation of the spiral (e.g., \cite{Ren2020}). Another promising diagnostic is the so-called GI wiggle, a characteristic oscillatory perturbation in the velocity field produced by gravitational instability \cite{Longarini2021, Terry2022} whose kinematic signature has recently been used to claim the detection of GI-driven spirals in AB Aurigae, a disk that exhibits a range of properties compatible with a massive self-gravitating disk on the verge of fragmentation \cite{Speedie2024}. Finally, another way to distinguish the origin of spirals is their size: while the spiral arms in some systems have been attributed to planetary companions, other systems exhibit spiral features that are too large and too massive to be easily explained by the presence of gas giant planets, providing hints that DI may play a role in these cases. As an example, the large ($\approx1000$~AU, \cite{Cleeves2016}) disk around IM Lup has been observed with clear gaps and spiral structures (left panel of Fig.~\ref{fig:disk_mass}), which some have suggested could be indicative of gravitational instability (e.g., \cite{Huang2018}). The disk mass has been measured dynamically to be $0.1~M_\odot$, with $Q\sim1$ in the outer regions of the disk (\cite{Lodato2023}, right panel of Fig.~\ref{fig:disk_mass}). These observations suggest that the conditions for fragmentation, such as disk mass, temperature, and cooling efficiency, may be met in some protoplanetary disks, and that DI could lead to the formation of planets in these environments. Furthermore, synthetic observations of disk fragmentation simulations suggest that the direct detection of clumps formed by GI is difficult but possible with ALMA in dust continuum at short wavelengths provided that the disk midplane can be directly observed  \cite{Mayer16} , yet this might be significantly more challenging in reality if a residual envelope, and eventually accreting filaments, surround the midplane.\\

\subsection{Direct imaging of mature systems}

\begin{figure}[h!]
   \centering
    \includegraphics[width=8.4cm]{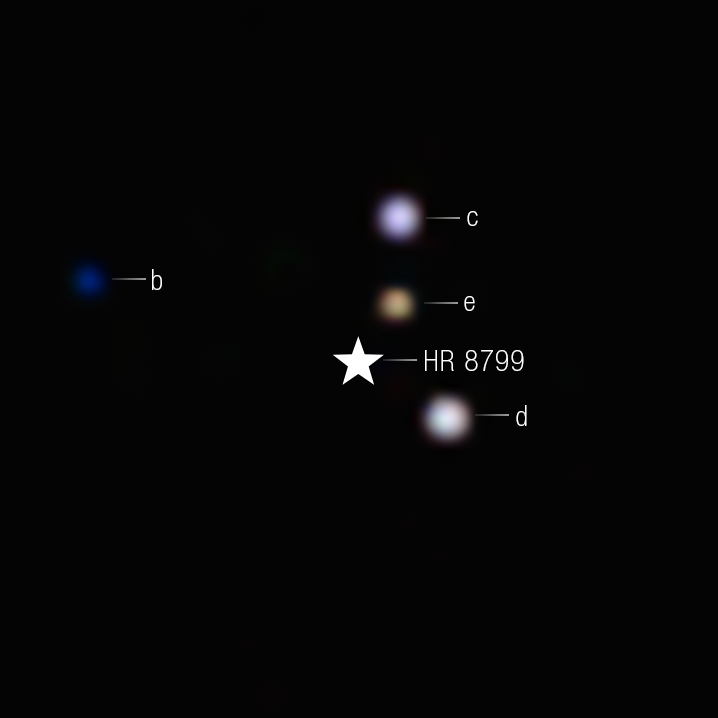}
    \caption[err]{
    {\small The HR 8799 system. A star symbol marks the location of the host star. The signals named b through e indicate the four giant planets. Colors are applied to filters from the JWST/NIRCam instrument, revealing their intrinsic differences. 
    Image from James Webb Space Telescope (NASA, ESA, CSA, STScI, Laurent Pueyo (STScI), William Balmer (JHU), Marshall Perrin (STScI)). 
    }
    \label{fig:hr8799}
}
\end{figure}

Direct imaging not only provides the opportunity to search for protoplanets in gravitationally unstable disks, but also to  investigate the planetary demographics  resulting by DI. As an example, the planetary system around HR 8977 is composed by four gas giant companions of several Jupiter masses each with semi-major axes $\lesssim70$ AU (see Fig.~\ref{fig:hr8799}, \cite{Balmer2025_HR8799}), and is often used to support the validity of the DI model. Recently, \cite{Nielsen2019} and \cite{Vigan2021} compared the companions occurrence rates of the GPI and SHINE surveys with populations expected from DI \cite{Forgan2018} and core accretion (CA) models \cite{Emsenhuber2021}. \cite{Nielsen2019} found that brown dwarfs and giant planets follow different distributions, with the former likely formed via DI, while the latter formed through CA. \cite{Vigan2021} compared their companion population with synthetic populations from both CA and DI. For FGK stars, they found that neither mechanism alone can explain the observed demographics, and both CA and DI are required, with CA likely playing the dominant role.

These conclusions were inferred on the basis of planetary masses obtained by converting photometric measurements through isochrones. However, these methods were unconstrained and subject to major uncertainties, namely the age estimate for the system, the post-formation conditions of the planets like hot-start vs. cold-start \cite{Marley2007, Mordasini2013,Marleau2014}, the assumed evolutionary and atmospheric models (e.g., cloudy vs.~clear). 

Recently, thanks to the microarcsecond astrometric precision provided by the VLTI/GRAVITY instrument and the complementarity of radial velocity and GAIA astrometric measurements, dynamical masses for a handful of directly imaged exoplanets could be measured. Dynamical masses are measured based on the orbital motion and are independent from prior assumptions like the system's age or the planet's initial conditions. So far only a fraction of the known imaged exoplanets have a measured dynamical mass, due to their often large separation and the poor coverage of their orbit. Among the most successful dynamical mass measurements we can list $\beta$ Pic b and c, AF Lep b and HR8799e \cite{Brandt2021_bPic, Brandt2021, Bonse2025, Balmer2025}. In addition to these relatively young planets, the first old ($>1$~Gyr) planet has been imaged with JWST/MIRI \cite{Matthews2024}, and thanks to ongoing RV monitoring its mass has been constrained to $6.3\pm0.6~M_J$. Eps Indi Ab will be instrumental in constraining and testing the mass-luminosity relationship for planets at old ages, and already showed some conflicts with current modeling of the planet atmosphere. Similarly, the Roman Microlensing Survey which is the microlensing component of the Roman Space Telescope's Galactic Bulge Time-Domain Survey is expected to detect hundreds of exoplanets on wide orbits as well as free-floating ones \cite{Boss2025}. Assumption-independent mass measurement for the population of directly imaged planets will be instrumental to calibrate the population synthesis models and assess whether these planets form indeed through gravitational instability.

\section{Summary \& Outlook}

A better understanding of the disk instability model requires improvements in both the theoretical (numerical simulations) and observational fronts. Despite many theoretical, numerical and observational efforts, key questions related to the DI model remain open. These include: 
\begin{itemize}
    \item Under what conditions do protoplanetary disks fragment and form clumps?
    \item What is the role of magnetic fields in clump formation? 
    \item What is the expected population of objects formed by DI? That is, what are the expected properties of these clumps? These include frequencies, masses, and compositions as well as final radial distances.
    \item How do clumps evolve over time? How does the evolution depend on their physical properties (mass, composition)? 
    \item What are the observational signatures that can be used to further constrain and validate the DI model for giant planet formation?  
\end{itemize}

Numerically, simulating gravitational fragmentation in protoplanetary disks requires high-resolution 3D (magneto)hydrodynamics with accurate radiative cooling prescriptions, high-resolution and longer simulation times. In addition, in order to properly simulate the long-term evolution of clumps,  processes such as mass accretion, core formation, and interactions with the disk and other clumps must be considered. 
Future advancements in radiation-hydrodynamic simulations and magnetohydrodynamic effects are important in improving our understanding of fragmentation and subsequent clump evolution. In addition, such numerical and theoretical results can be fed into global DI models which combine many of the governing physical effects and conduct planetary population syntheses, providing a quantitative link to observations.
\par 

Observationally, distinguishing disk instability from core accretion remains challenging due to degeneracies in planetary properties. ALMA observations of massive, cold disks have provided indirect evidence for gravitational instability, but direct detections of forming clumps remain elusive. While direct imaging of wide-orbit gas giants, such as the HR 8799 and PDS 70 systems, suggests possible gravitational collapse at large separations, robust population synthesis models struggle to reconcile observed planet distributions with efficient fragmentation \cite{Vigan2021}. Additionally, the metallicities of exoplanet atmospheres, expected to be lower in disk instability scenarios than in core accretion, remain uncertain due to observational limitations. Future prospects include JWST spectroscopy of young exoplanets, ELT high-contrast imaging, and more detailed ALMA surveys of young disks, which will provide better constraints on the frequency and properties of gravitationally unstable regions, helping to assess the viability of disk instability as a dominant formation mechanism.
The DI model may operate in parallel with core accretion, and a comprehensive understanding of giant planet formation likely requires a hybrid approach that incorporates both mechanisms, depending on the disk environment and its properties.
With growing observational power and advancing simulations, the DI model can illuminate some of the most elusive stages of planet and star formation. We are looking forward to the years ahead, and believe that the DI model will become a vital framework, guiding us closer to unraveling the mysteries of disk evolution and planetary formation.

%\section*{Appendix}
%\addcontentsline{toc}{section}{Appendix}
%
%

%\input{references}
\bibliographystyle{author/spmpsci}
\bibliography{references}
\end{document}